\documentclass[english,10pt,twoside]{article}
\usepackage{amsmath}
\usepackage[english]{babel}
\usepackage{makeidx}
\usepackage{natbib}
\usepackage{pdflscape}
\usepackage{graphicx}
\usepackage{caption}
\usepackage{subcaption}
\usepackage{siunitx}
\usepackage{geometry}
\newcommand{\bleriot}  {Bl\'{e}riot}
\newcommand{\aap}      {Astronomy and Astrophysics\ }
\newcommand{\aj}       {Astronomical Journal\ }

\newcommand{\apjl}     {Astrophysical Journal Letters\ }
\newcommand{\icarus}   {Icarus}
\newcommand{\mnras}    {Monthly Notices Royal Astron. Soc.\ }
\newcommand{\planss}   {Planetary and Space Science\ }
\newcommand{\nat}      {Nature\ }
\newcommand{\science}      {Science\ }

\begin{document}

\title{A Librational Model for the Propeller \bleriot{} in the 
Saturnian Ring System}
\author{M. Seiler *, M. Sei{\ss}, H. Hoffmann and F. Spahn\\{
  \small Theoretical Physics Group, Institute of Physics and Astronomy, 
  University of Potsdam}\\
  {\small * miseiler@uni-potsdam.de}}
\maketitle

\begin{abstract}
The reconstruction of the orbital evolution of the propeller structure 
\bleriot{} orbiting in Saturn's A ring from recurrent observations in 
Cassini ISS images yielded a considerable offset motion from the expected 
Keplerian orbit \citep{Tiscareno2010ApJL}. 
This offset motion can be composed by three sinusoidal harmonics with 
amplitudes and periods of 1845, 152, 58 km and 11.1, 3.7 and 2.2 years, 
respectively \citep{Sremcevic2014EPSC}. 
In this paper we present results from N-Body simulations, where we  
integrated the orbital evolution of a moonlet, which is placed at the 
radial position of \bleriot{} under the gravitational action of the 
Saturnian satellites. Our simulations yield, that especially the gravitational 
interactions with Prometheus, Pandora and Mimas is forcing the moonlet to
librate with the right frequencies, but the libration-amplitudes are far too 
small to  explain the observations. Thus, further mechanisms are needed to 
explain the amplitudes of the forced librations -- e.g. moonlet-ring 
interactions. Here, we develop a model, where the moonlet is allowed to be 
slightly displaced with respect to its created gaps, resulting in a breaking 
point-symmetry and in a repulsive force. As a result, the evolution of the 
moonlet's longitude can be described by a harmonic oscillator. In the presence 
of external forcings by the outer moons, the libration amplitude gets the more 
amplified, the more the forcing frequency gets close to the eigenfrequency of 
the disturbed propeller oscillator. Applying our model to \bleriot{}, it is 
possible to reproduce a libration period of 13 years with an amplitude 
of about \SI{2000}{\kilo\meter}. 
\end{abstract}

\section{Introduction}
One of the most puzzling discoveries of the spacecraft Cassini -- orbiting 
around Saturn since its arrival in June 2004 -- has been the observation 
of disk-embedded moons orbiting within Saturn's main rings 
\citep{Tiscareno2006Nat,Spahn2006Nat, Sremcevic2007Nat, Tiscareno2008AJ}.
The typical density variations downstream the embedded moonlet's orbit are 
created by the gravitational interaction of a small sub-kilometer-sized object 
(called \textit{moonlet}) with the surrounding ring material and reminds of a 
two-bladed \textit{propeller} giving the structure its name 
\citep{Spahn2000AAP, Sremcevic2002MNRAS}. Meanwhile more than 150 propeller
structures have been detected within the A and B ring. The largest propeller
structure which is a few thousands km in azimuth is called \bleriot{} and is 
caused by a moonlet with a diameter of around 800 meters.
However, the moonlet is still too small to allow its direct observation by the 
Cameras aboard the spacecraft Cassini.

Nevertheless, the propeller structure permits the observation and orbital
tracking of the largest moonlets. The reconstruction of the orbital evolution
of \bleriot{} revealed an offset motion with respect to a Keplerian motion of
considerable amplitude \citep{Tiscareno2010ApJL}. \citet{Tiscareno2010ApJL}
found, that one possibility to describe the longitudinal excess motion of
\bleriot{} is a harmonic function of 300 km and a period of 3.6 years.

It is still an ongoing debate, whether \bleriot{}

\begin{itemize}
\item[i)] is librating due to gravitational interactions with the other moons
in the Saturnian system (resonances),
\item [ii)] is suffering from stochastical interactions with the surrounding
ring material \citep{Rein2010AAP, Tiscareno2013PSS},
\item[iii)] is in a 'frog resonance' \citep{Pan2010ApJL,Pan2012ApJ},
\item[iv)] or if it is even perturbed by the combined effects of all the above.
\end{itemize}

Considering hypothesis (ii), several attempts, have been started in order
to explain this excess motion \citep{Pan2010ApJL,Pan2012ApJ, Tiscareno2013PSS},
where \citet{Rein2010AAP} delivered an explanation considering a stochastic
migration and \citet{Pan2012MNRAS} showed in N-Body simulations, that such a
mechanism could generate an maximal excess motion of 300 km over a time of 4
years.

Another mechanism (iii) has been introduced with a so-called "frog
resonance" model, where \citet{Pan2010ApJL} consider a resonance between the 
moonlet and its created gap edges, being modeled as two co-orbital point masses.
This interaction is causing the moonlet to librate within its gap. However, the
approximation of the gaps as coorbital point masses of almost half of the mass
of the moonlet is a strong simplification, which does not reflect the true
structure of a propeller.

Meanwhile newer orbital fits of the orbital evolution of \bleriot{} have
been performed, tracking and reconstructing the orbit from a larger set of ISS
images over a larger time span. The most recent investigation by
\citet{Sremcevic2014EPSC} yielded, that \bleriot's orbital excess motion can
be fitted astonishingly well by three harmonic functions with amplitudes
and periods of 1845, 152 and 58 km and 11.1, 3.7 and 2.2 years, respectively,
where the standard deviation of the remaining residual is about 17 km
\citep[see also][]{Spahn2017Book}.

The harmonic behavior of the excess motion might suggest that resonant
interactions with the other Saturnian moons serve as a reason for the excess
motion.

Such harmonic systematic deviations from the expected Keplerian orbit are a
known phenomenon in the Saturnian system. Some of the outer moons are librating
systematically, like the moon Enceladus which is in a 2:1 ILR with Dione and
the moon Atlas, which is perturbed by the 54:53 CER and 54:53 ILR by
Prometheus \citep{Goldreich1965cMNRAS, Goldreich2003aIcar,Goldreich2003bIcar,
Spitale2006AJ,Cooper2015AJ}.

Following these examples, we perform simulations where the moonlet is perturbed
by the moons of Saturn to characterize the orbital motion of \bleriot.
It will turn out, that this approach can explain the observed frequencies, but
not the large libration amplitudes. Thus, we propose a model of ring-moonlet
interactions which is capable to explain the amplification of the perturbed 
excess motion to observable excursions. 
Favoring hypothesises (i) and (iv) outer gravitational near resonant 
excitations and ring-moonlet interactions are needed to explain the 
observations.

The paper is organized as follows: In section \ref{sec:integration} we
present the N-Body integrations and their results. In section
\ref{sec:toy-model} we introduce our model of the moonlet-gap
interaction and give a connection to the azimuthal motion in section
\ref{sec:toy-model-relation}, and then apply our model to the moonlet
Bl\'{e}riot. Finally, we will conclude and discuss our results
in section \ref{sec:conclusion-discussion}.

\section{Test Moonlet Integration} \label{sec:integration}

For the numerical integrations we consider the gravity of 15 Saturnian moons 
of masses $m_i$ and the oblate (up to $J_6$) Saturn of mass $m_c$ and radius
$R_c$ determining the dynamics of the moonlet

\begin{align}
\ddot{\vec{r}} & = \nabla \left\{ \frac{Gm_c}{r} \left[1 - \sum_{j=2}^\infty J_j 
  \left( \frac{R_c}{r}\right)^j P_j \left(\cos \vartheta \right) \right] 
  - \sum_i^N G m_i \left( \frac{1}{\left|\vec{r}_i - \vec{r} \right|} - \frac{ 
                 \vec{r}_i \cdot \vec{r}}{r_i^3} \right) \right\} \, .
\end{align}

The considered moons are: Atlas, Daphnis, Dione, Enceladus, Epimetheus, Hyperion,
Iapetus, Janus, Mimas, Pan, Pandora, Prometheus, Rhea, Tethys and Titan, where 
the initial values for the 15 moons were taken from the SPICE kernels sat375.bsp 
and sat378.bsp at the initial time 2000-001T12:00:00.000.

Next, we apply our N-Body integration routine to the propeller structure 
\bleriot, which we model as a test particle, placed in the 
Saturnian equatorial plane ($i_0=0$ deg) on a circular orbit ($e_0 = 0$) at 
the expected orbital position \citep[initial semi-major axis 
$a_0 = 134912.125 \, \text{km}$ and mean longitude 
$\lambda_0 = 259.45 \, \text{deg}$, private communication][extrapolation 
of results from ISS images for the initial time given above
]{Sremcevic2013private}. 
 
\begin{figure*}
  \centering
  \includegraphics[width=0.49\textwidth]{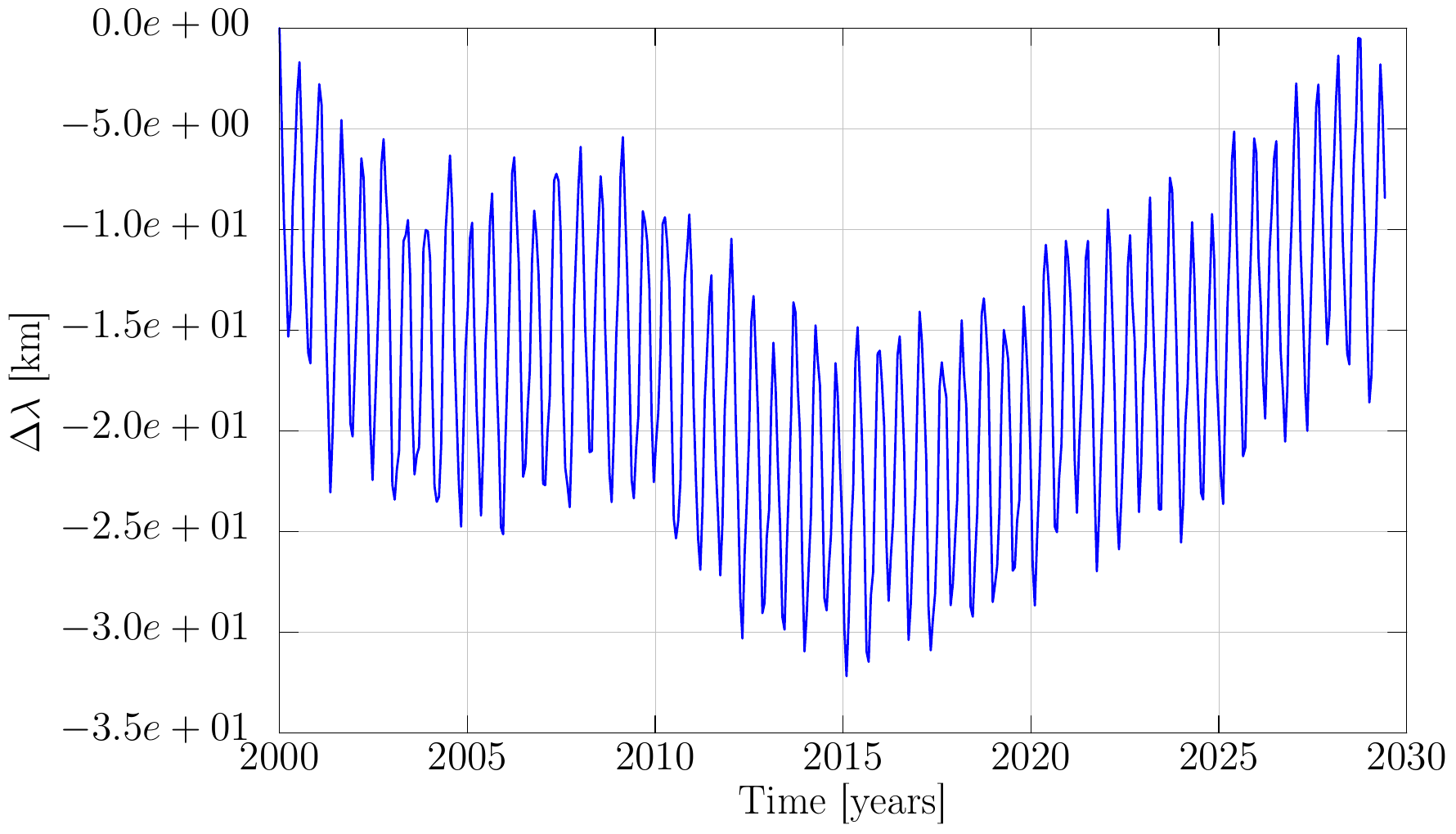}
  \includegraphics[width=0.49\textwidth]{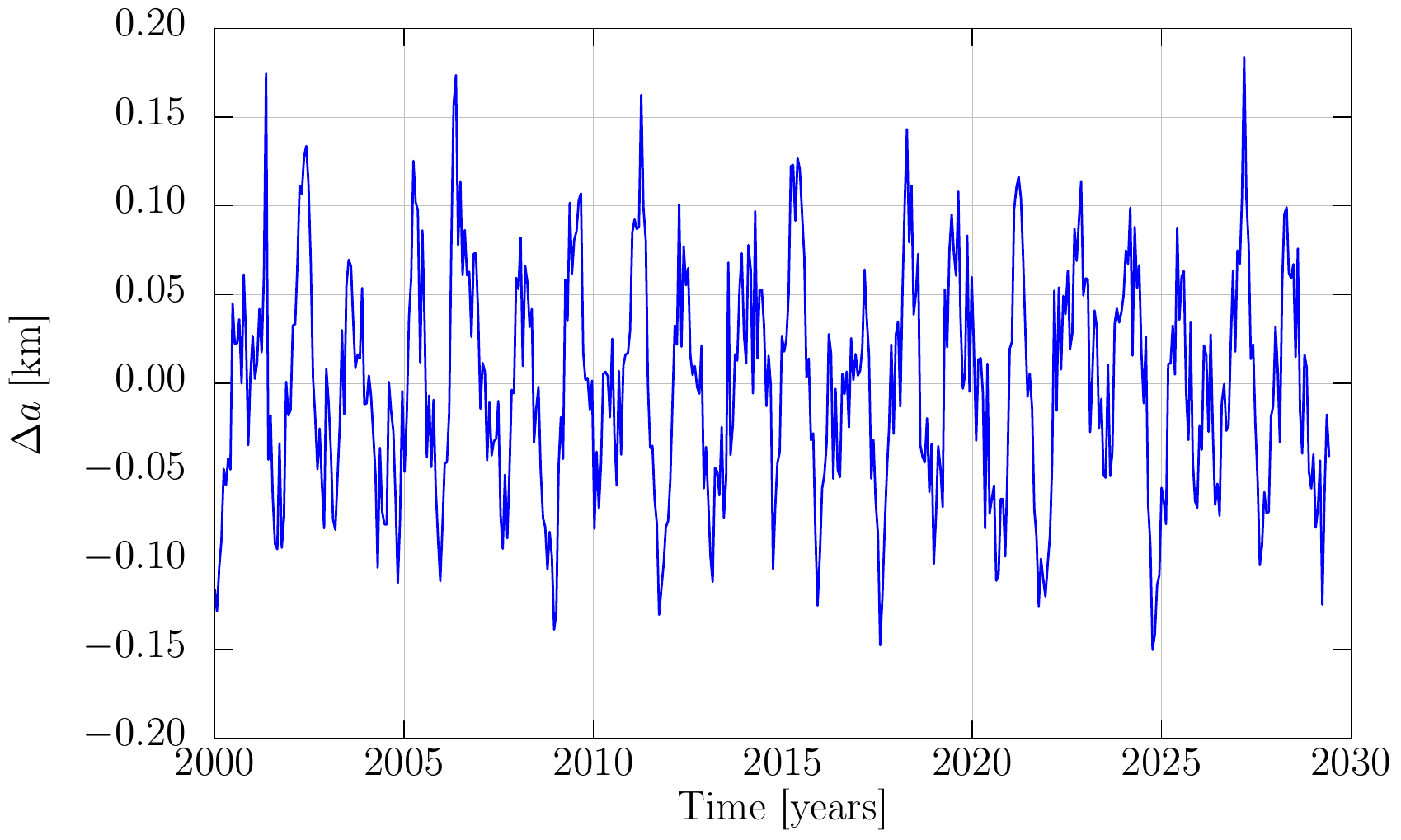}
  \includegraphics[width=0.49\textwidth]{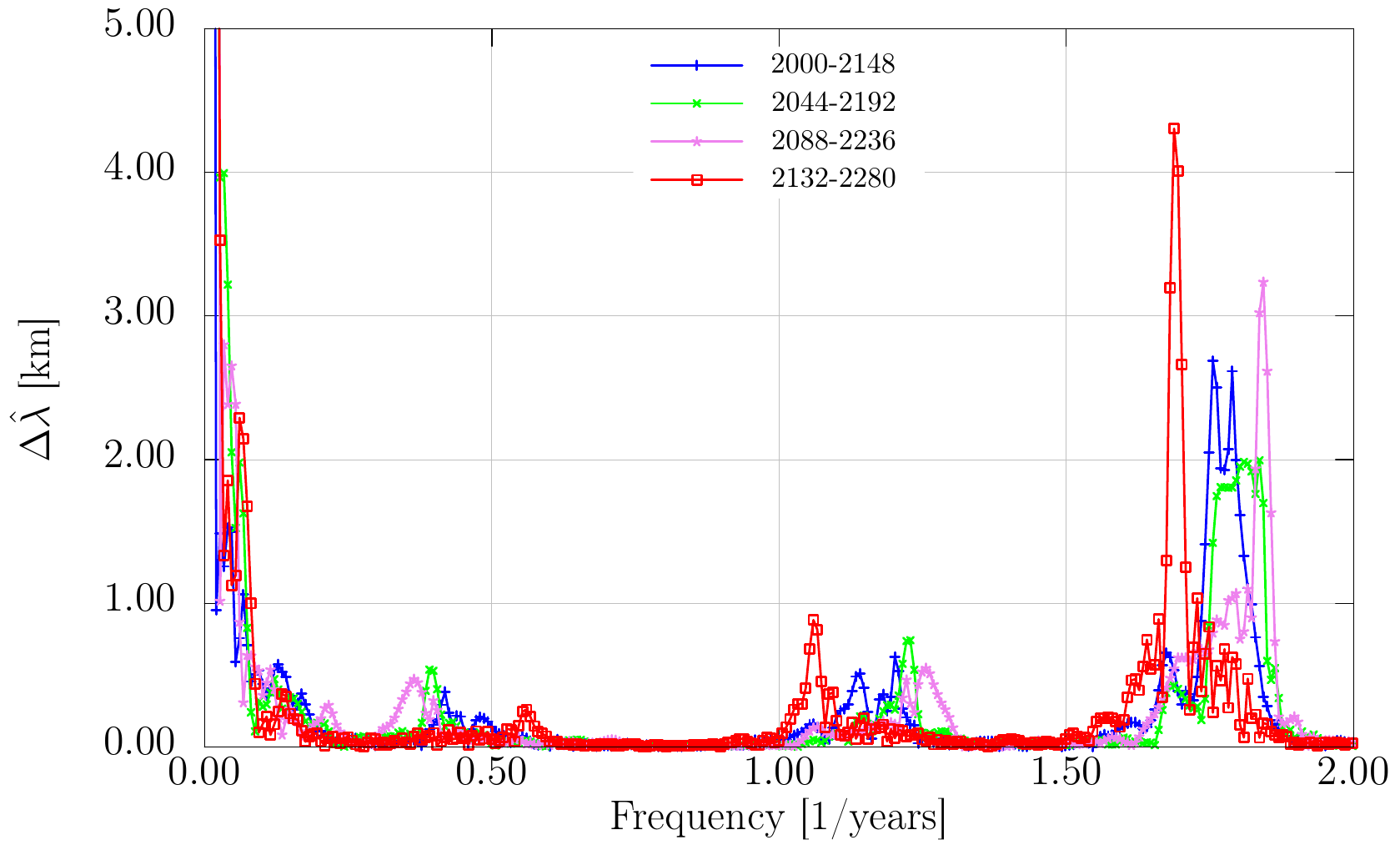}
  \includegraphics[width=0.49\textwidth]{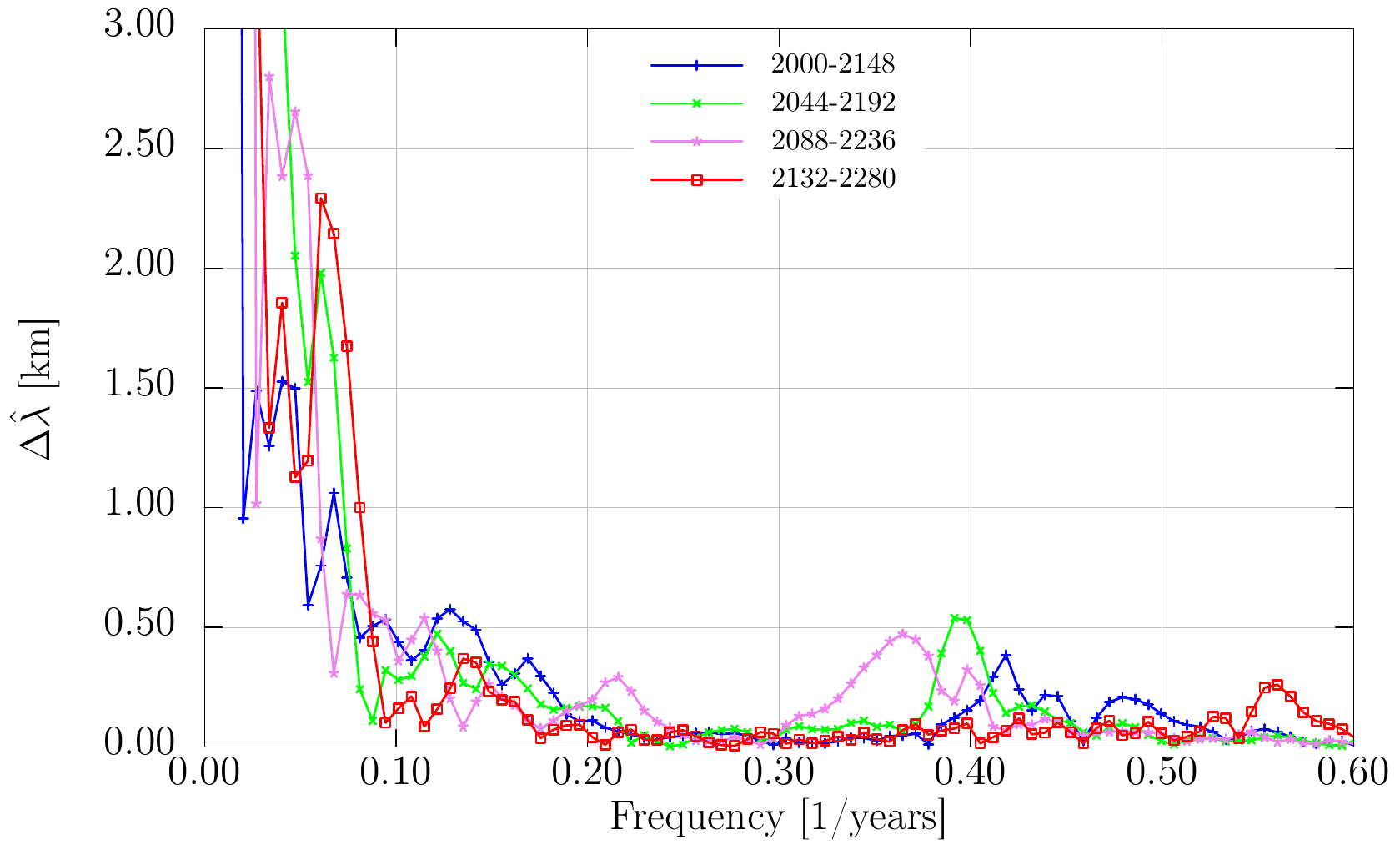}
  \caption{\footnotesize Evolution of the orbital elements of a  
    test particle, placed at the orbital position of Bl\'{e}riot 
    ($a_0 = 134912.125$ km, $\lambda_0 = 259.45$ deg, $e_0 = 0$ and 
    $i_0 = 0$ deg). 
    Top Left: Mean longitude residual for the time span from 2000 to 2030, 
    where a linear trend with mean motion of $n_{av} = 616.7819374$ deg 
    day$^{-1}$ has been subtracted. Top Right: Residual in semi-major axis for
    the same time span, with $a_{av} = 134912.24$ km. 
    Bottom Left: Fourier spectrum of the mean longitude residual of  
    \bleriot{} from a 300 Earth years simulation run. The simulation data has 
    been divided into 4 frames of 148 Earth years length and an overlay of  
    70\%. It is clearly visible, that the Fourier spectrum is not stationary and 
    thus its periods and amplitudes are changing with time. 
    Bottom Right: Detailed view of Fourier spectrum in the frequency range 0 to
    0.6 years$^{-1}$ of \bleriot{} from the left panel. 
    As already visible from the top left panel, the half year 
    libration is the most dominant oscillation, being a result of the 14:13 
    resonance with Pandora. More libration periods can be found and are listed 
    in Table \ref{tab:ble-periods}.}
  \label{fig:ble-geoelem}
\end{figure*}

The numerical deviation of the mean longitude and the semi-major axis of 
\bleriot{} over a time span of 30 years are shown in the upper panels of 
Figure \ref{fig:ble-geoelem}.

Analyzing the frequency spectrum (see in the lower panels) a clear librational 
behavior with different frequencies and amplitudes is visible in the residuals of the mean longitude and semi-major axis. The gravity of the acting 15 moons 
induce a mean eccentricity of $e_{av} = 2.9 \cdot 10^{-6}$ and inclination of 
$i_{av} = 7.5 \cdot 10^{-5}$ deg on the test moonlet. 

\begin{table*}
  \centering
  \begin{tabular}{|c|c|c|c|c|c|} \hline
    $f$ [years$^{-1}$] & $T$ [years] & $\Delta \hat{\lambda}$ [km] & $\Delta
    \hat{n}$ [m s$^{-1}$] & $\Delta \hat{a}$ [m] & Resonance  \\ \hline \hline
   1.7 & 0.6 &  4.6  &  1.5 $\cdot$ 10$^{-3}$ & 8  & 14:13 CER Pnd-Ble  \\ \hline
   1.25 & 0.8 &  1.0  &  2.2 $\cdot$ 10$^{-4}$ & 1.2  & 3:16:13 Mim-Pnd-Ble
   (2)                         \\ \hline
    0.4 & 2.5 &  0.15 &  4.3 $\cdot$ 10$^{-5}$ & 0.5 & 3:16:13 Mim-Pnd-Ble
   (1)                          \\ \hline
   0.07 & 14.3 & 2.3  &  3 $\cdot$ 10$^{-5}$ & 0.15  & 3:16:13 Mim-Pnd-Ble
   (3)                          \\ \hline
  \end{tabular}                                                                 
  \caption{\footnotesize 
    Libration periods and their amplitudes in the mean 
    longitude estimated from the Fourier spectral analysis of the simulation data. 
    Additionally, the amplitudes in mean motion, and semi-major axis and the
    corresponding resonance are given. 
    Additionally to the 14:13 CER of Pandora also three-body-resonances between
    Mimas, Pandora and \bleriot{} seem to play an important role.  }
  \label{tab:ble-periods}
\end{table*}

\subsection{The Dominating Moons}

\begin{figure*}
\centering
\includegraphics[width=0.49\textwidth]{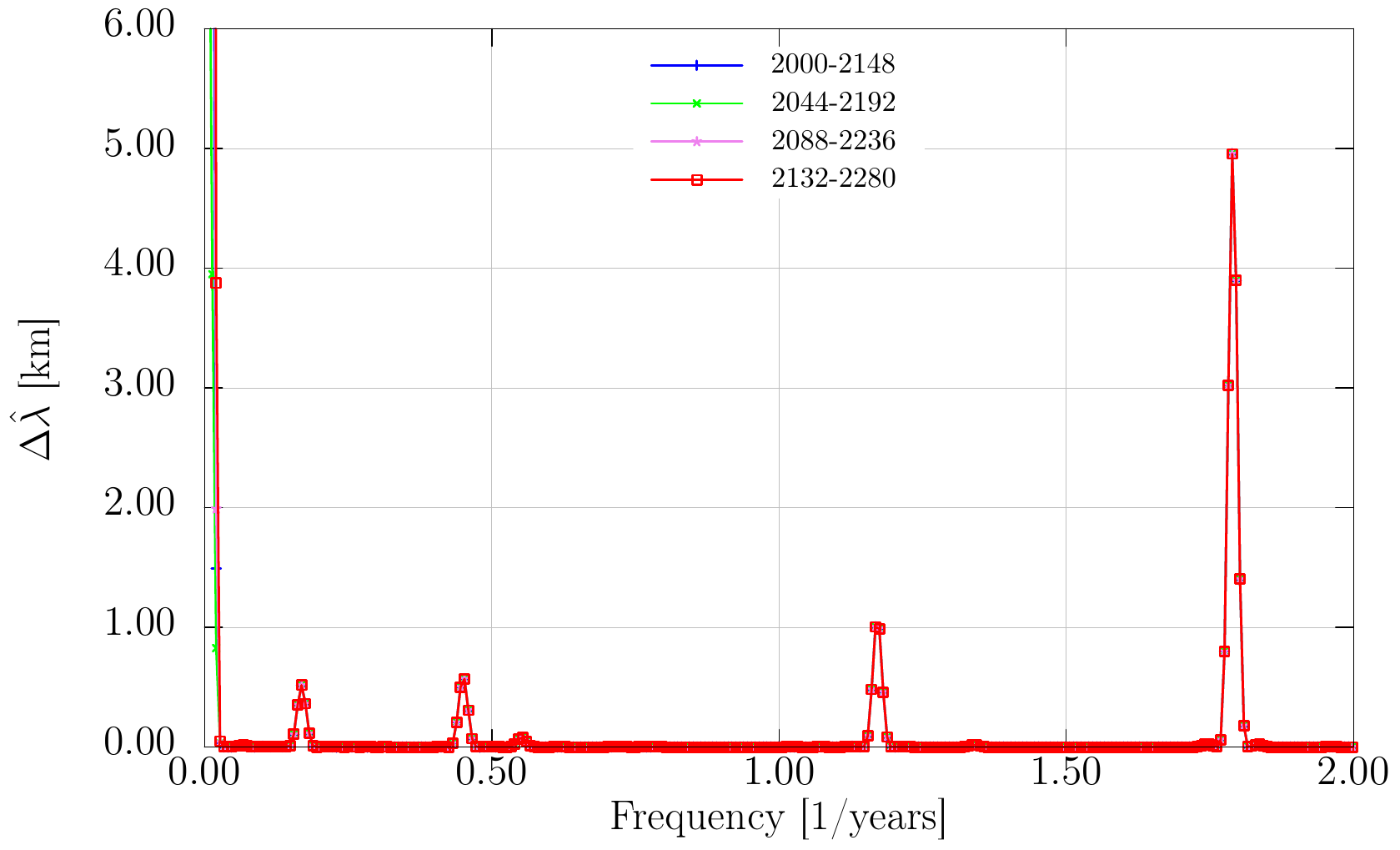}
\includegraphics[width=0.49\textwidth]{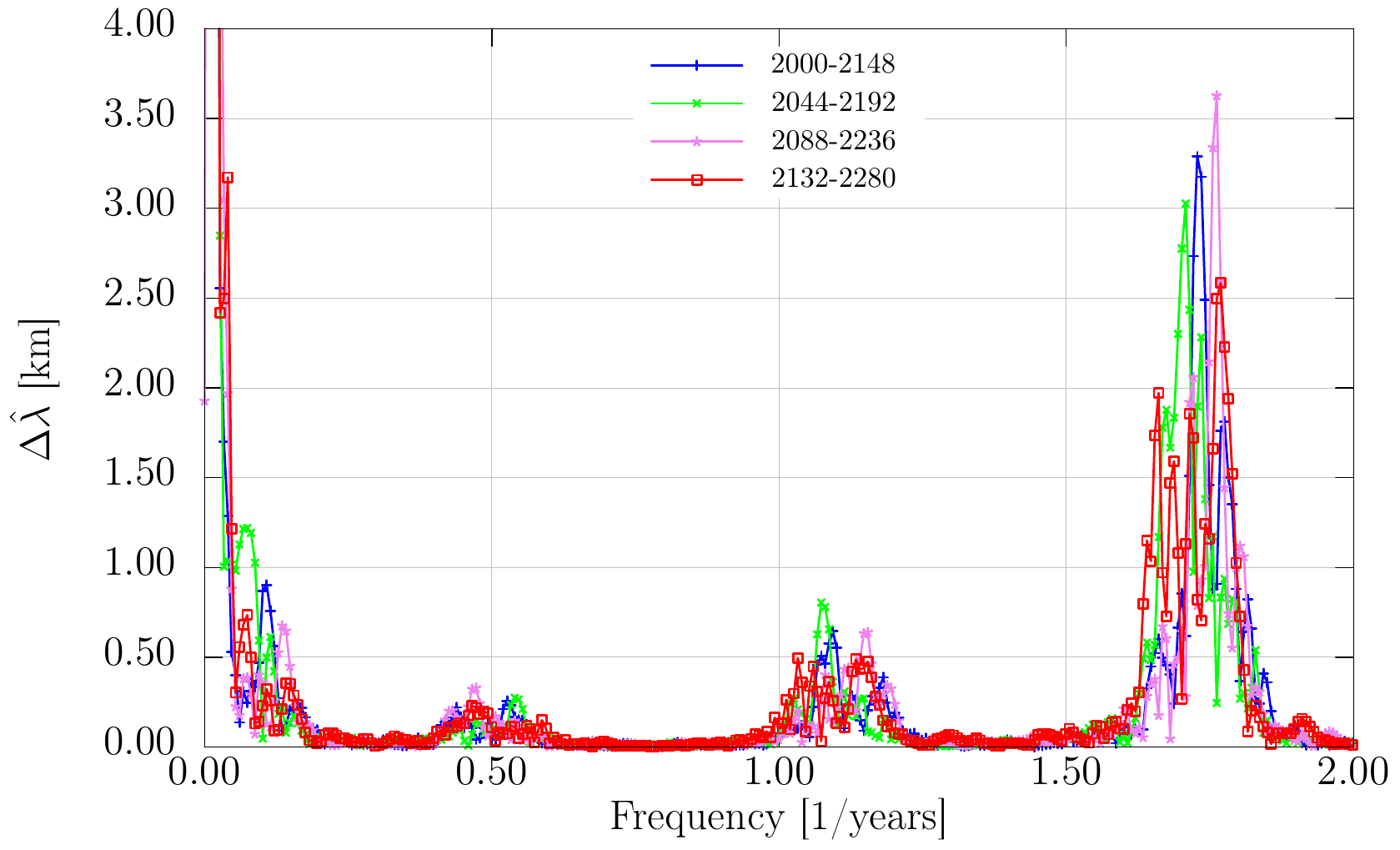}
\caption{\footnotesize 
  Fourier spectral analyses to find the dominating moons perturbing the orbit of 
  \bleriot. Left: Amplitude spectrum of the mean longitude 
  considering only Pandora and Mimas. Mimas and Pandora are in a 3:2 resonance 
  and additionally form a 3:16:13 three body resonance with Bl\'{e}riot. 
  Right: Spectrum, considering Prometheus, Pandora, Mimas, Tethys and Titan. 
  The gravity of Prometheus, Titan and Tethys is influencing the orbital 
  evolution of the complete system. Additionally, the gravity of Prometheus 
  results in a non-stationary signal in the Fourier spectrum. Thus, the 
  libration frequencies are changing due to the interaction between Pandora and 
  Prometheus.}
\label{fig:dom-moons-ble}
\end{figure*}

The amount of moons has been decreased systematically in order to identify the 
ones, which cause considerable resonant perturbations on \bleriot. 
It turns out, that the satellites Pandora and Mimas are dominating the resonant 
behavior, resulting in four characteristic peaks in the Fourier spectrum of 
the mean longitude residual (see Figure \ref{fig:dom-moons-ble}, left panel). 

The most dominating influence is clearly found in the
14:13 corotation-eccentricity resonance (short CER) of Pandora, which is forcing 
\bleriot{} to librate with a period of 0.6 years with an amplitude of 5 km. A
complete list of the important periods, amplitudes and resonances,
causing the librations can be found in Table \ref{tab:ble-periods}.

The presence of Mimas is resulting in a three-body resonance between Mimas, 
Pandora and \bleriot{}. Mimas and Pandora are known to be in a 3:2 resonance,
resulting in a libration period of 1.8 years. Additionally, Pandora is
perturbing the orbital evolution of \bleriot{} with its 14:13 resonance.
Considering the resonant arguments \citep{Murray1999Book} of both 
resonances (3:2 and 14:13) one can construct related three-body resonances 
with libration periods of 0.6, 0.8, 2.5 and 14.3 years (compare with Figure 
\ref{fig:dom-moons-ble} and Table \ref{tab:ble-periods}) by subtracting the
corresponding resonant arguments:

\begin{align}
  \varphi_{3:16:13,1}  & = 3 \lambda_{Mim} - 16 \lambda_{Pnd} + 
  13 \lambda_{Ble}   \\
  \varphi_{3:16:13,2}  & = 3 \lambda_{Mim} - 16 \lambda_{Pnd} + 
  13 \lambda_{Ble} - \varpi_{Mim} + \varpi_{Pnd}   \\
  \varphi_{3:16:13,3}  & = 3 \lambda_{Mim} - 16 \lambda_{Pnd} + 
  13 \lambda_{Ble} - \varpi_{Pnd} + \varpi_{Ble}   \\
  \varphi_{3:16:13,4}  & = 3 \lambda_{Mim} - 16 \lambda_{Pnd} + 
  13 \lambda_{Ble} - \varpi_{Mim} + \varpi_{Ble} \, .
\end{align}

Adding Prometheus, Titan and Tethys, all having influence on the orbital
dynamics of Pandora and Mimas, leads to a non-stationary signal in the
Fourier spectrum. This could be caused by the chaotic and strong interactions
between Prometheus and Pandora resulting in wandering libration frequencies
and changing amplitudes in the orbital dynamics of \bleriot. Performing
long-time simulations for the test moonlet considering a time span of up to 100
Saturnian years, the 42:40 IVR of Prometheus seems to become more and more
important forcing the moonlet to librate with a period of about 4 years with a
radial amplitude of 20 meters and an amplitude around 1 km in mean longitude.
This signal is clearly visible in the Fourier spectrum and stationary in
contrast to the other signals.

Although the libration periods of the test moonlet from our simulations agree
fairly well with the observational data, the resulting amplitudes are far too
small.

\section{Moonlet-Propeller Interactions} \label{sec:toy-model}

Next, we will consider the gravitational interaction between the embedded
moonlet with its created gap region. Imagine a non-symmetric propeller
structure so that the moonlet gets accelerated by the ring gravity.
We will show, that the evolution of the related longitudinal residual can be 
described by a harmonic oscillator. This harmonic oscillator feels a periodic 
external forcing caused by the gravity of the outer moons. The amplitudes of 
the external forced frequencies get the more magnified, the 
closer the forcing frequency matches the eigenfrequency of the harmonic 
oscillator (propeller-moonlet system).

\begin{figure}
  \centering
  \includegraphics[width=0.4\textwidth]{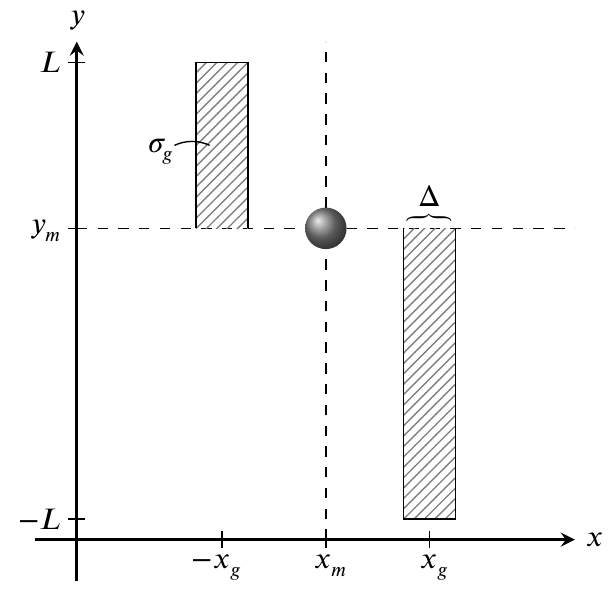}
  \caption{
    Sketch of our propeller toy model, illustrating the symmetry 
    breaking of the propeller lobes. The central moonlet is located at position 
    $(x_m,y_m)$, defining the beginning of the gap structures in this way. 
    The partial gaps, created by the moonlet-disc interactions are 
    located at $x=x_g$ and $x=-x_g$ with lengths $L \pm y_m$ and width 
    $\Delta$.  }
  \label{fig:propeller_model}
\end{figure}

Consider a moonlet located at ($x_m,y_m$) as illustrated in Figure
\ref{fig:propeller_model}. Interacting with the surrounding viscous
ring material, the propeller moonlet gravitationally scatters ring particles
to larger and smaller orbits creating two gaps in its vicinity in the ring
material (regions of reduced surface mass density) and which are decorated by two
density enhanced regions pairwise downstream of its orbit \citep{Spahn1989Icar}.
Viscous diffusion of the ring material counteracts this gap-creation,
smoothing out the structure with growing azimuthal distance downstream the
moonlet \citep{Spahn2000AAP, Sremcevic2002MNRAS}. For simplicity, we
assume the gap shape to be a rectangular area with reduced density $\sigma_g$,
illustrated by the shaded regions at $\left|x\right| = x_g$ and width $\Delta$ in
Figure \ref{fig:propeller_model}. The diffusion process defines the
length $L \pm y_m$ of the gaps, while the width of the propeller-blades is set
by the moonlet's mass $m$ or its Hill radius $h$

\begin{equation}
\frac{\Delta}{2} \propto h(m) \sim a_0 \left( \frac{m}{3m_c} \right)^{\frac{1}{3}}
\, ,
\end{equation}  

and Saturn's mass is labeled by $m_c$. We assume that both gaps are 
anchored at the position $y_m$ following the moonlet's motion. Further, the
ends at $L$ and $-L$ are fixed, because the imprint of the moonlet motion will 
be averaged out along the azimuth due to the diffusion process and dragged along
with the Kepler shear. 

In the symmetric case ($x_m=0,y_m=0$) the torque on the moonlet due to the
gravitational interaction with the gap is zero because of point symmetry.
When the moonlet is leaving its mean position, the gravitational 
force $\vec{F}_g$ of the ring material on the moonlet is

\begin{equation}
  \vec{F}_g \left(\vec{r}_m\right) = -G \int_{-\infty}^{\infty} \, dx \, 
  \int_{-\infty}^{\infty} \, dy \; \sigma\left(\vec{r}\right) \frac{\vec{r}_m-
  \vec{r}}{\left|\vec{r}_m-\vec{r}\right|^3} \; ,
  \label{equ:force}
\end{equation}

where $\sigma(\vec{r}), \vec{r} = (x,y)^T$ and $\vec{r}_m = (x_m,y_m)^T$ are 
the surface mass density and the position vectors of the ring material and of 
the moonlet.

Splitting the integrals, subtracting the unperturbed background density
$\sigma_0$ and evaluating the integrals up to first order in $x_m$ and $y_m$, 
the azimuthal force acting on the moonlet reads

\begin{align}
  F_{g,y} \left(x_m,y_m\right) & \approx 2G\left|\delta\sigma\right| \Delta 
  \left( \frac{x_m}{x_g^2} + \frac{y_m}{L^2} \right) \ \,  ,
  \label{equ:Fy_final}
\end{align}

where $\delta\sigma = \sigma_g - \sigma_0$. 
In the following we assume that the azimuthal excursions of the moonlet are 
negligibly small compared to the azimuthal extent of the propeller structure 
($y_m \ll L$). We also assume that the radial displacement of the moonlet is 
very small compared to the radial distance of the gaps to the moonlet 
($x_m \ll x_g$).

\subsection{Relation to the Azimuthal Libration} \label{sec:toy-model-relation}

The mean motion is given by the change in the mean longitude

\[ n = \frac{d \lambda}{d t} . \]

Thus, the change in azimuthal direction of the moonlet is given by the Gaussian 
perturbation equation

\begin{equation}
  \frac{d n}{d t} = \frac{d^2 \lambda}{d t^2} \approx - \frac{3}{a} F_y \, ,
\label{equ:omega}
\end{equation}

neglecting higher orders in eccentricity, where we used the relation 
$\lambda = y_m a^{-1}$. Inserting $F_{g,y}$ from Eq. (\ref{equ:Fy_final}) 
results in 

\begin{equation}
  \frac{d^2 \lambda}{d t^2} = \frac{6}{a} G \delta\sigma 
  \Delta \frac{y_m}{L^2} = 6G \delta\sigma \frac{\Delta}{L^2} 
  \lambda = - \omega^2 \lambda \, ,
\label{equ:period}
\end{equation}

with $\omega^2 = 6G \left|\delta\sigma\right| \frac{\Delta}{L^2}$ the libration 
frequency and $T = 2\pi \omega^{-1}$ the libration period. 
In this lowest order of the perturbation expansion we arrive at a harmonic
oscillator. Here, the eigenfrequency contains the properties of the propeller
feature as the lengths $\Delta$ and $L$, but also the drop in the surface mass 
density $\delta \sigma$.

\subsection{Application to \bleriot{}}

In order to estimate the eigenlibration period for the moonlet, we use values,
which have been inspired by theoretical and observational data
\citep{Tiscareno2007Icar, Spahn2000AAP}.

We can estimate the length of the gaps from the analytical model derived by
\citet{Sremcevic2002MNRAS}. The relative mass moved out of the gaps along the
azimuth is

\[ \int_0^L \frac{\sigma_g - \sigma_0}{\sigma_0} \, dy = -2.1 aK \, , \]

which can be obtained from a direct numerical integration.
Assuming $\delta \sigma / \sigma_0 = -0.5$ for the rectangular simplification of
the gaps, the gap length is $L=4.2 aK$ and thus, for Bl\'{e}riot $L \approx 600
km$, where the scaling length $aK$ can be calculated by

\[ aK = 160 km \left[ \frac{400 \frac{cm^2}{s}}{\nu}\right] \left[\frac{h}{500
  m}\right]^3 \left[\frac{\Omega}{10^{-4} \frac{1}{s}}\right] \, \]

  with $\nu, h$ and $\Omega$ denoting the viscosity, the Hill radius and the
  Kepler frequency. The value of $\nu$ fits to extrapolated values from
  observations of \citet{Tiscareno2007Icar} and simulations of
  \citet{Daisaka2001Icar}.

  Further, the width of the gap $\Delta = 2 h \approx 1 km$.

  As a result, we estimate a eigenlibration period of that ring-moonlet harmonic
  oscillator:

  \[ T = \frac{2\pi L}{\sqrt{6G \left|\delta\sigma\right|\Delta}}
    \approx 13 yr \left[ \frac{L}{600 km} \left(
      \frac{\left|\delta\sigma\right|}{20 \frac{g}{cm^2}} \frac{\Delta}{1 km}
        \right)^{-\frac{1}{2}} \right] . \]

The Hill radius $h$ of the propeller moonlet has a strong influence on the 
gap length $L\propto h^3 \sim m$ and thus on the libration period. With 
$h = 480m \pm 50m$ the libration period is $T = (13 \pm 4)\text{yr}$. 

If the propeller moonlet is placed in the Saturnian ring system with all the 
other moons, this harmonic oscillator will be forced with a zoo of frequencies
due to the Saturnian moons. 
When a forcing frequency matches the eigenfrequency close enough, a large gain 
in the resulting amplitude is possible. For example, if the period of the 
resonant interaction and the eigenperiod of the oscillator differ by a third 
of a year, the amplitude is amplified by two orders of magnitude. The closer 
the forcing frequency is to the eigenfrequency, the larger the resulting 
amplitudes. Formally, to get to the observed amplitudes, which are about three 
orders of magnitude larger than the values from Table \ref{tab:ble-periods}, 
the difference of the periods has to be smaller than $0.03$ years.
The longest period from Table \ref{tab:ble-periods} falls well into the 
uncertainty interval of the estimated libration period $T$, so that our model 
can indeed induce much larger libration amplitudes. 

One should note, that our simple model is generally not fully consistent, 
because the libration amplitude $y_m$ is not much smaller than the used gap 
length $L$. Thus, the condition to approximate the force for small amplitudes 
is easily violated. However, in reality the gap extents for more than several
thousand km and $L$ is rather a decaying length of the gap if one for example 
assumes a exponential relaxation of the gap. A more complex model, would remove 
this inconsistency, but the oscillatory behavior due to the repulsive azimuthal 
force should persist. We will address such a comprehensive non-linear model in
the future.

\section{Conclusion and Discussion} \label{sec:conclusion-discussion}

In this paper we have presented results of N-Body simulations, which
characterize the orbital evolution of the moonlet of the propeller
\bleriot, being perturbed by the gravitational interactions with 15 outer
and inner moons of Saturn. We found, that the 14:13 CER of Pandora and the
3:16:13 three-body resonances of Mimas and Pandora
are the dominating perturbations of \bleriot's orbit.
Additionally, the chaotic interaction between Pandora and Prometheus has an
effect on \bleriot's orbital evolution as well, resulting in changing libration
frequencies over time.

Our simulations yielded, that the gravitational interactions with the other
moons cause \textit{similar libration frequencies} of \bleriot{} as concluded
from the Cassini ISS images, but the corresponding amplitudes are too small.
Considering propeller-moonlet interactions with a new model, we have been able
to find a mechanism to amplify certain modes in form of a harmonic oscillator
which is periodically driven by an external forcing.

The Eigenfrequency of this oscillating system contains key-properties of the
propeller structure. In the presence of an external forcing, this
oscillating system amplifies the induced forced oscillations by the outer
satellites up to several orders of magnitude. The closer the forcing frequency
is
to the eigenfrequency the stronger the amplification can be. Applying our
moonlet-propeller-gap interaction model to \bleriot, reasonable results have
been obtained for the libration period, which fit the observations fairly well.
Combining our model with the simulation results, we are able to reproduce the
largest observed mode for the libration of \bleriot. The smaller observed
modes are not reproduced, but may evolve if one regards processes of non-linear
mode coupling in our model, which we did not consider yet.

Even this first simplifying theoretical investigation demonstrates, that outer
gravitational perturbations in combination with ring-moonlet interactions are 
necessary to address the propeller-moonlet migration problem.

In order to verify our results, we plan to incorporate our model into
hydrodynamical simulations in the future. Additionally, we want to add higher 
order terms in the evaluation, i.e. nonlinear terms, in order to study mode 
coupling effects.

Besides the improvements on the propeller model, we also plan to apply our
N-Body integrations to the other propeller structures located in the outer
A ring.

\end{document}